\documentstyle[camera,prbplug,psfig]{article}

\begin{document}

\title{ {\small JOURNAL OF THE PHYSICAL SOCIETY OF JAPAN 70 (2001) 2833}\\
\vspace{0.5cm}
Anisotropic optical conductivity of Nd$_{2-x}$Ce$_x$CuO$_4$ thin films
}

\author{Takashi Yanagisawa$^1$, Shigeru Koikegami$^1$, Hajime Shibata$^1$, 
Shinji Kimura$^2$, 
Satoshi Kashiwaya$^1$, Akihito Sawa$^3$, Noritaka Matsubara$^{1,4}$
and Koki Takita$^4$}

\address{$^1$Nanoelectronics Research Institute, National Institute of 
Advanced Industrial Science and Technology Central 2 1-1-1 Umezono, 
Tsukuba, Ibaraki 305-8568, Japan\\
$^2$National Metrological Laboratory, National Institute of Advanced Industrial
Science and Technology Central 3, 1-1-1 Umezono, Tsukuba, Ibaraki 305-8568,
Japan\\
$^3$Correlated Electron Research Center, National Institute of 
Advanced Industrial Science and Technology Central 4 1-1-1 Umezono, 
Tsukuba, Ibaraki 305-8568, Japan\\
$^4$Tsukuba University, 1-1-1 Tennoudai, Tsukuba, Ibaraki 305-8577, Japan}

\maketitle

\begin{abstract}
Optical conductivity spectra $\sigma_1(\omega)$ of Nd$_{2-x}$Ce$_x$CuO$_4$ 
thin films, measured by the reflectance-transmittance method (R-T method) 
which has been proposed to investigate the far-infrared 
(FIR) spectroscopy, are investigated based on the anisotropic pairing 
model.  Precise measurements of the frequency-dependent conductivity 
$\sigma_1(\omega)$ enable us to examine quantitatively the nature of 
the superconducting gap through FIR 
properties for the first time in high-$T_c$ superconductors.  We show that the 
behavior of optical conductivity $\sigma_1(\omega)$ is consistent with the
anisotropic superconducting gap and is well explained by the 
formula for $d$-wave pairing in the low-energy regime of the FIR region.  
Our results suggest 
that the electron-doped cuprate superconductors Nd$_{2-x}$Ce$_x$CuO$_4$ have 
nodes in the superconducting gap.
\end{abstract}

\section{Introduction}
Oxide high-$T_c$ superconductors have been investigated intensively over the
last decade.\cite{lt99,lld01}  $d$-wave superconductivity is now well 
established
for hole-doped superconductors.  However, there is a class of high-$T_c$
superconductors doped with electrons,\cite{tok89,tak89} for which both 
$s$-wave\cite{kas98} and $d$-wave pairing\cite{tsu00,kok00,pro00} have been 
reported.
Nd$_{2-x}$Ce$_x$CuO$_4$ is a typical example
of electron-doped materials and the symmetry of Cooper pairs has been a
controversial issue.
It is important to examine the symmetry of Cooper pairs in the study of
high-$T_c$ superconductors.

Far-infrared (FIR) spectroscopy is a powerful technique to investigate the 
nature of the superconducting gap. The conventional FIR spectroscopy based on
a Kramers-Kronig transformation, however,
is rather unfavorable for studying electronic properties in the low energy
regime of the FIR region.\cite{tim89,gao93}
Since the superconducting gap $\Delta$ in 
Nd$_{2-x}$Ce$_x$CuO$_4$
is very small, there have been no reports on the study of the nature of the
superconducting gap of Nd$_{2-x}$Ce$_x$CuO$_4$ through such techniques,
although there have been a number of reports on the FIR spectroscopy of
Nd$_{2-x}$Ce$_x$CuO$_4$.\cite{hom97,ono99}
Recently, a new method to examine FIR spectroscopy has been
developed without the need for evaluating the Kramers-Kronig 
transformation.\cite{kog98}
In this method, the optical conductivity is estimated from the data of 
reflectance spectra $R(\omega)$ and transmittance spectra $T(\omega)$ by
substituting them into a set of coupled equations.  The new method is free
from the conventional difficulties in the FIR region since we do not
need the aid of the Kramers-Kronig transformation.\cite{shi01}
This method is referred to as the R-T method in this paper.
 
The purpose of this paper is to investigate FIR optical properties of 
Nd$_{2-x}$Ce$_x$CuO$_4$ obtained by the R-T method from the viewpoint of 
unconventional
superconductors.  We will show that the available data for the optical
conductivity and transmittance are well explained by the
$d$-wave pairing model in the clean limit.  The value of the superconducting 
gap is estimated as $2\Delta\sim 50-60$ cm$^{-1}$, which is consistent with the
available value estimated by scanning tunneling spectroscopy.\cite{kas98}

\begin{figure}
\centerline{\psfig{figure=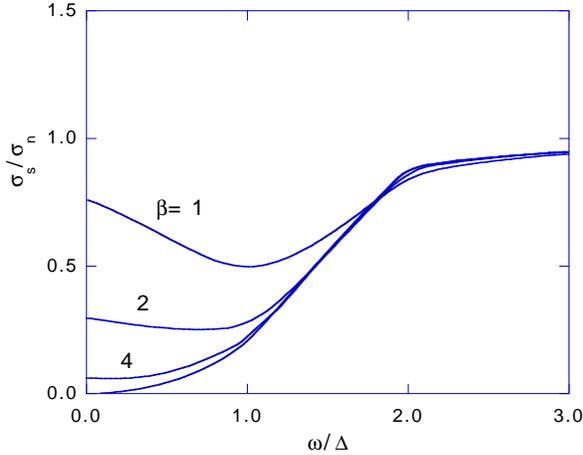,width=10cm}}
\caption{
Optical conductivity as a function of $\omega$ for several values of
temperature.  From the top $T/\Delta=1/\beta=1,1/2,1/4$ and 0.
}
\label{fig1}
\end{figure}

\begin{figure}
\centerline{\psfig{figure=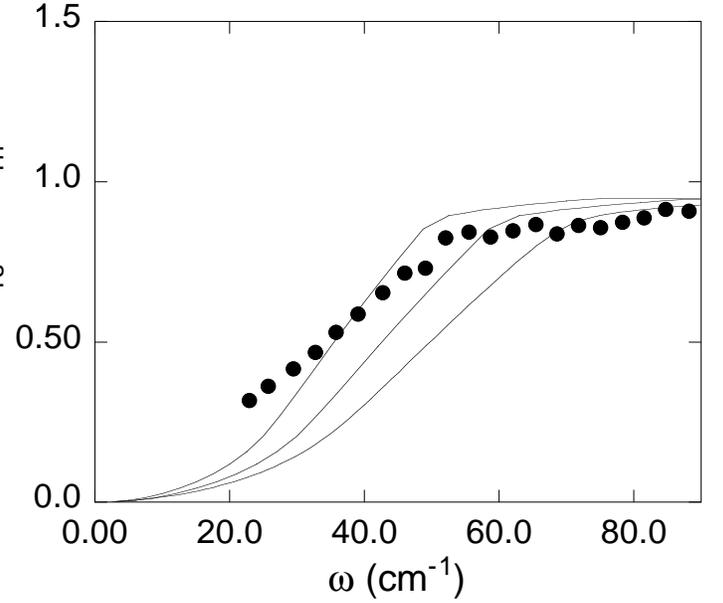,width=10cm}}
\caption{
Optical conductivity (circles) by the R-T method and
theoretical predictions at $T=0$ (solid curves).
From the left $2\Delta=50$cm$^{-1}$, 60cm$^{-1}$ and 70cm$^{-1}$.
}
\label{fig2}
\end{figure}

\section{Theory and Measured Conductivity}
The frequency-dependent conductivity $\sigma(\omega)$ has been calculated by
Mattis and Bardeen,\cite{mat58} Abrikosov {\em et al.}\cite{abr59} and 
Skalski {\em et al.}\cite{ska64} for
isotropic superconductors.  The original Mattis-Bardeen theory was tested
for a conventional type-I $s$-wave superconductor, where the coherence
length $\xi$ and magnetic penetration depth $\lambda$ satisfy $\xi\gg\lambda$.
The opposite limit $\xi\ll\lambda$ (London limit) was also examined for 
$s$-wave
pairing by field theoretical treatments.\cite{ska64}
For the high-$T_c$ compounds of type-II superconductor with small coherence 
length, 
the formula in the London limit is appropriate for optical conductivity
measurements.  Recently, the conductivity $\sigma(\omega)$ of an
unconventional superconductor has been examined theoretically in the London 
limit.\cite{hir89,hir92,hir94,gra95,dor00}
The current response function, from which the optical conductivity is derived,
is given 
by\cite{hir89}

\begin{equation}
K_{ij}({\bf q},\omega_m)= -\frac{e^2k_F^2}{m^2c}\sum_k \hat{k}_i\hat{k}_j
\frac{1}{\beta}\sum_n Tr[G({\bf k}_+,\epsilon_n+\omega_m)G({\bf k}_-,\epsilon_n)],
\end{equation}
where ${\bf k}_{\pm}={\bf k}\pm {\bf q}/2$ and $\epsilon_n=(2n+1)\pi/\beta$.
The single-particle matrix Green's function is given by
\begin{equation}
G({\bf k},\epsilon_n)= \frac{i(\epsilon_n-\Sigma(\epsilon_n))\tau^0+\xi_k\tau^3+\Delta_k\tau^1}{(\epsilon_n-\Sigma(\epsilon_n))^2+\xi_k^2+\Delta_k^2},
\end{equation}
where $\Delta_k$ is the anisotropic order parameter and $\Sigma(\epsilon_n)$ 
is the
self-energy due to impurity scattering. 
$\tau^i$ ($i=0,1,\cdots$) denote Pauli matrices.
Since we consider the case where $\xi\ll\lambda$ holds, the real part of
optical conductivity is well approximated by the formula in the London limit:
\begin{equation}
\sigma_{ij}(\omega)= -\frac{c}{\omega}\lim_{q\rightarrow 0}{\rm Im}
K_{ij}({\bf q},\omega).
\end{equation}
The expressions of $K_{ij}({\bf q},\omega)$ for real and imaginary parts were 
given in ref. 20.
Our focus is the collisionless limit of the normalized conductivity to
compare it with the data for Nd$_{2-x}$Ce$_x$CuO$_4$ since $\xi\ll\ell$ holds
for the mean-free path $\ell$.
For anisotropic superconducting order parameter $\Delta_k$ such that 
the average over the Fermi surface vanishes $\langle\Delta_k\rangle=0$,
the expression for $\sigma_{xx}(\omega)$ in the collisionless limit on the 
plane is simply given by\cite{hir92}
\onecolumn
\begin{equation}
\frac{\sigma_{xx}(\omega)}{\sigma_n}= \frac{1}{2\omega}
\int_{-\infty}^{\infty}dx\langle{\rm Re}
\frac{|x|}{(x^2-\Delta_k^2)^{1/2}}\rangle
\langle{\rm Re}\frac{|x-\omega|}{[(x-\omega)^2-\Delta_k^2]^{1/2}}\rangle
[{\rm tanh}(\frac{\beta x}{2})-{\rm tanh}(\frac{\beta(x-\omega)}{2})],
\end{equation}
\twocolumn
which is an angle-dependent generalization of the Mattis-Bardeen formula.
For the $d$-wave symmetry, the average over the Fermi surface denoted by the
angular brackets is defined as
\onecolumn
\begin{equation}
\langle{\rm Re}\frac{x}{(x^2-\Delta_k^2)^{1/2}}\rangle=
{\rm Re} \int \frac{d\phi}{2\pi}\frac{x}{[x^2-(\Delta{\rm cos}(2\phi))^2]^{1/2}},
\end{equation}
\twocolumn
where the order parameter is factorized as $\Delta_k=\Delta{\rm cos}(2\phi)$.
In Fig.1, we show the behaviors of $\sigma_s(\omega)\equiv \sigma_{xx}(\omega)$
as a function of $\omega$ for several values of temperature $T$.
The FIR behaviors reflect the lines of nodes on the Fermi surface.

FIR reflection $R(\omega)$ and transmission $T(\omega)$ measurements were 
performed for
Nd$_{2-x}$Ce$_x$CuO$_4$ ($x=0.15$) thin films deposited by laser ablation 
onto (001)  MgO substrates.  The thickness of the Nd$_{2-x}$Ce$_x$CuO$_4$ 
thin film was about 40 nm.
$T_c$ was estimated to be $\sim$20K.  The electric field
of the FIR radiation was predominantly parallel to the $a$-$b$ plane.
The conductivity spectra were evaluated by the R-T method from the data
for $R(\omega)$ and $T(\omega)$ at $T=4.3$ and 30K.\cite{shi01}

The R-T method provides us with reliable spectroscopic data in the FIR 
region due to which a comparison between the experimental data and that of 
theoretical
analysis is possible.  In the R-T method, both the refrectance spectra 
$R(\omega)$ 
and the transmittance spectra $T(\omega)$ are measured experimentally from
which a set of coupled equations is derived describing the transmittance and
reflectance of a thin film on a substrate.  The coupled equations are solved
numerically by the Newton method to determine the optical conductivity.
This method is free from the difficulties in the FIR region which occur
commonly in the conventional method employing a Kramers-Kronig transformation. 
In Fig.2, we show the observed data and theoretical curves at $T=0$ for
$2\Delta=50$, 60 and 70 cm$^{-1}$.  The experimental data 
$\sigma_1(3.4K)/\sigma_1(30K)$ normalized by the normal state values at $T=30$K 
are shown in Fig.2.  It is obvious from the experimental results that there 
is no evidence of a true gap, which is suggestive of an anisotropic 
superconducting gap, since the spectral weight of conductivity should vanish for 
$\omega\le 2\Delta$ at $T=0$ in conventional isotropic superconductors.
It is also shown in Fig.2 that they are well fitted by the curve with 
$2\Delta=50$ cm$^{-1}$, which is consistent with the value estimated by
scanning tunneling spectroscopy measurements.\cite{kas98}

A transmission curve is also presented in Fig.3, where $T_S/T_N$, the ratio of
the transmission in the superconducting ($T=$3.4K) to that in the normal state 
($T=$30K), is the
experimentally measured quantity.  The following phenomenological expression 
for $T_S/T_N$
is employed to determine the transmission curve theoretically,\cite{glo57}
\onecolumn
\begin{equation}
\frac{T_S}{T_N}= \frac{1}{[T_N^{1/2}+(1-T_N^{1/2})(\sigma_1/\sigma_n)]^2+
[(1-T_N^{1/2})(\sigma_2/\sigma_n)]^2},
\end{equation}
\twocolumn
where $\sigma_1$ and $\sigma_2$ are real and imaginary parts of the
conductivity $-(c/\omega) K({\bf q},\omega)$ for ${\bf q}\rightarrow 0$, respectively.  
$T_N$ is determined as $T_N\simeq 0.08$ from the
expression for the ratio of the power transmitted with a film to that with
no film given as
\begin{equation}
T_N= 1/[1+\sigma_n d\frac{Z_0}{n+1}]^2 .
\end{equation}
Here, $d$ is the film thickness, $n$ is the index of refraction of the 
substrate,
and $Z_0$ is the impedance of free space.  We have assigned the following
values;  $d=4\times 10^{-6}$cm, $n=3.13$, $Z_0=377 \Omega$,
$\sigma_n\approx 7\times 10^3 \Omega^{-1}$cm$^{-1}$.
Apparently, the $\omega$-dependence of measured transmittance agrees with the 
theoretical curve for
$2\Delta=50\sim$60cm$^{-1}$ as shown in Fig.3.
The agreement between the observed quantities and theoretical curve is
significant, which should be compared to the isotropic BCS prediction
calculated from the Mattis-Bardeen equations.\cite{cho96}

\begin{figure}
\centerline{\psfig{figure=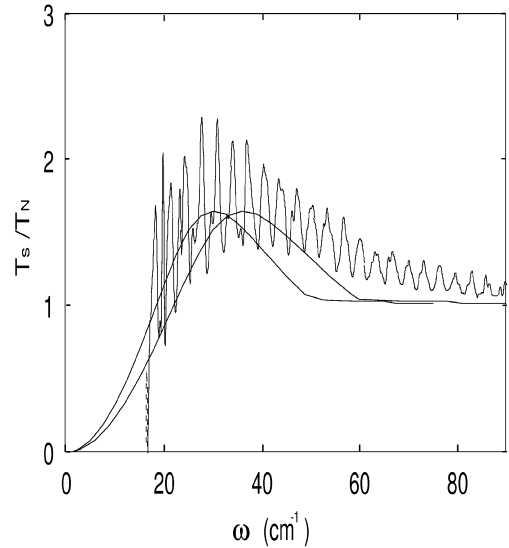,width=8cm}}
\caption{
Observed transmittance and the 
theoretical curve at $T=0$ (solid curve) for $2\Delta=50$cm$^{-1}$ (left)
and 60cm$^{-1}$ (right).
}
\label{fig3}
\end{figure}

\section{Summary}
We have successfully made a comparison between experimental data and 
theoretical data
for the optical conductivity of Nd$_{2-x}$Ce$_x$CuO$_4$ in the FIR 
region for the first time.
We have shown that there is a reasonable agreement between the optical
conductivity $\sigma_1(\omega)$ observed by the R-T method and theoretical analysis without
adjustable parameters except the superconducting gap.  An estimate of
50$\sim$60 cm$^{-1}$ for the superconducting gap is consistent from both
the experimental and theoretical aspects.
The FIR optical conductivity suggests that the superconducting gap
of electron-doped Nd$_{2-x}$Ce$_x$CuO$_4$ is an unconventional one with nodes
on the Fermi surface.
The anisotropic nature of electron-doped superconductors is consistent
with the recent researches performed for the one-band and three-band Hubbard
models.\cite{yam98,yan00,yan01,kur99}
If the superconducting gap is anisotropic for the electron-doped
superconductors, there is a possibility that both the hole-doped and 
electron-doped cuprates superconductors are governed by the same 
superconductivity mechanism.

\end{document}